\documentclass[preprintnumbers,article,amsmath,amssymb,floatfix,10pt,prd,twocolumn,
superscriptaddress,nofootinbib]{revtex4-2}
\usepackage{bm}
\usepackage{amsfonts}
\usepackage{latexsym}
\usepackage[latin1]{inputenc}
\usepackage{graphicx}
\usepackage{amsmath}
\usepackage{palatino}
\usepackage{mathpazo}
\usepackage{textcomp}
\usepackage{float}
\usepackage{booktabs}
\usepackage{dcolumn}
\usepackage{ragged2e}
\usepackage{hyperref}
\hypersetup{colorlinks,citecolor=blue}
\hypersetup{colorlinks=true,linkcolor=red,filecolor=magenta,    urlcolor=blue}
\usepackage{amsmath}
\usepackage{xcolor}
\usepackage{orcidlink}
\usepackage{epsfig}
\usepackage{caption}
\usepackage{subcaption}
\usepackage{commath}
\def\jnl@style{\it}
\def\aaref@jnl#1{{\jnl@style#1}}

\def\aaref@jnl#1{{\jnl@style#1}}

\def\aj{\aaref@jnl{AJ}}                   % Astronomical Journal
\def\apj{\aaref@jnl{ApJ}}                 % Astrophysical Journal
\def\apjl{\aaref@jnl{ApJ}}                % Astrophysical Journal, Letters
\def\apjs{\aaref@jnl{ApJS}}               % Astrophysical Journal, Supplement
\def\apss{\aaref@jnl{Ap\&SS}}             % Astrophysics and Space Science
\def\aap{\aaref@jnl{A\&A}}                % Astronomy and Astrophysics
\def\aapr{\aaref@jnl{A\&A~Rev.}}          % Astronomy and Astrophysics Reviews
\def\aaps{\aaref@jnl{A\&AS}}              % Astronomy and Astrophysics, Supplement
\def\mnras{\aaref@jnl{Mon.~Not.~Roy.~Astron.~Soc.}}             % Monthly Notices of the RAS
\def\prd{\aaref@jnl{Phys.~Rev.~D}}        % Physical Review D
\def\prc{\aaref@jnl{Phys.~Rev.~C}}  % Physical Review C
\def\prl{\aaref@jnl{Phys.~Rev.~Lett.}}    % Physical Review Letters
\def\qjras{\aaref@jnl{QJRAS}}             % Quarterly Journal of the RAS
\def\skytel{\aaref@jnl{S\&T}}             % Sky and Telescope
\def\ssr{\aaref@jnl{Space~Sci.~Rev.}}     % Space Science Reviews
\def\zap{\aaref@jnl{ZAp}}                 % Zeitschrift fuer Astrophysik
\def\nat{\aaref@jnl{Nature}}              % Nature
\def\aplett{\aaref@jnl{Astrophys.~Lett.}} % Astrophysics Letters
\def\apspr{\aaref@jnl{Astrophys.~Space~Phys.~Res.}} % Astrophysics Space Physics Research
\def\physrep{\aaref@jnl{Phys.~Rep.}}      % Physics Reports
\def\physscr{\aaref@jnl{Phys.~Scr}}       % Physica Scripta
\def\commat{\aaref@jnl{Comm.~Math.~Phys.}}              % Communications in Mathematical Physics
\def\science{\aaref@jnl{Science}}               % Science
\def\cqg{\aaref@jnl{Classical Quant.~Grav.}}            % Classical and Quantum Gravity
\def\jpcs{\aaref@jnl{JPCS}}                                     % Journal of Physics Conference Series
\def\ijmpd{\aaref@jnl{Int.~J.~Mod.~Phys.~D}}                    % International Journal of Modern Physics D
\def\grg{\aaref@jnl{Gen.~Relat.~Gravit.}}               % General Relativity and Gravitation
\def\rpp{\aaref@jnl{Rep.~Prog.~Phys.}}          % Reports on Progress in Physics
\def\npa{\aaref@jnl{Nucl.~Phys.~A}}        % Nuclear Physics A
\def\lrr{\aaref@jnl{Living Rev.~Rel.}}                   % Living reviews in relativity
\def\jcap{\aaref@jnl{J.~Cosmology Astropart.~Phys.}}    % Journal of cosmology and astroparticle physics
\def\rmp{\aaref@jnl{Rev.~Mod.~Phys.}}   %Reviews of modern physics
\def\epjc{\aaref@jnl{Eur.~Phys.~J.~C}} 
\def\plb{\aaref@jnl{~Phy.~Lett.~B}} 
\def\mpla{\aaref@jnl{Mod.~Phy.~Lett.~A}} 
\def\arxiv{\aaref@jnl{arxiv.org}}

\allowdisplaybreaks[1]
\begin{document}
%\color{red}
\color{black}       %% For one column
\title{Reconstruction of $\Lambda$CDM Universe in $f(Q)$ Gravity}
%\end{document}

\author{Gaurav N. Gadbail\orcidlink{0000-0003-0684-9702}}
\email{gauravgadbail6@gmail.com}
\affiliation{Department of Mathematics, Birla Institute of Technology and
Science-Pilani,\\ Hyderabad Campus, Hyderabad-500078, India.}

\author{Sanjay Mandal\orcidlink{0000-0003-2570-2335}}
\email{sanjaymandal960@gmail.com}
\affiliation{Department of Mathematics, Birla Institute of Technology and
Science-Pilani,\\ Hyderabad Campus, Hyderabad-500078, India.}

\author{P.K. Sahoo\orcidlink{0000-0003-2130-8832}}
\email{pksahoo@hyderabad.bits-pilani.ac.in}
\affiliation{Department of Mathematics, Birla Institute of Technology and
Science-Pilani,\\ Hyderabad Campus, Hyderabad-500078, India.}
%
%%%%%%%%%%%%%%%%%%%%%%%%%%%%%%%%%%%%%  DATE  %%%%%%%%%%%%%%%%%%%%%%%%%%%%%%%%%%%%

\begin{abstract}
 In this manuscript, we present a number of fascinating explicit reconstructions for the $f(Q)$ gravity from the background of Friedmann-La\^imatre-Robertson-Walker (FLRW) evolution history. We find the more general functions of non-metricity scalar $Q$ that admit exact $\Lambda$CDM expansion history. Adding extra degrees of freedom to the matter sector is the only method to get the scale factor to behave in this manner for more generic functions of $Q$. In addition, a cosmological reconstruction for modified $f(Q)$ gravity is constructed in terms of e-folding. It is shown how any FLRW cosmology can arise from a specific $f(Q)$ theory. We also reconstruct the well-known cosmological evolution for the specific examples of $\Lambda$CDM cosmology.

\end{abstract}

\maketitle
\date{\today}

%%%%%%%%%%%%%%%%%%%%%%%%%%%%%%%%%%%%%%%%%%%%%%%%%%%%%%%%%%%%%%%%%%%%%%%%
%%%%%%%%%%%%%%%        Introduction        %%%%%%%%%%%%%%%%%%%%%%%%%%%%%
%%%%%%%%%%%%%%%%%%%%%%%%%%%%%%%%%%%%%%%%%%%%%%%%%%%%%%%%%%%%%%%%%%%%%%%%
\section{Introduction}
For the last two decades, one prominent debatable problem has been the current acceleration of the universe. The recent advances in the observational cosmology confirm this phenomenon of the universe; for example the type Ia supernovae \cite{Perlmutter/1999, Riess/1998,Riess/2004}, Cosmic Microwave Background Radiation \cite{Spergel/2003} and large-scale structure observations \cite{Koivisto/2006,Daniel/2008,Nadathur/2020}. But, the real issue arises when it comes to the theoretical development to address this process of the universe. It is well-known that, after more than one hundred years, general relativity (GR) is still the most successful fundamental gravitational theory to describe the large-scale structure of the universe. When we apply it to cosmology, the FLRW spacetime and the matter source give some exact solutions for the scale factor $a(t)$, and it helps us to study the expansion of the Universe. If someone wishes to recover FLRW spacetime, then the universe needs to go through two types of accelerated expansion. One is the early time inflation, which dynamics can be studied by introducing the scalar field in the Einstein-Hilbert action. And the present accelerated expansion can be explained by adding a cosmological constant in the Einstein field equation. This is the very simple model (called $\Lambda$CDM model), which is able to explain the late-time cosmic acceleration of the universe and fit all the available observations discussed formerly. But unfortunately, this beautiful model is affected by fine-tuning problems related to the vacuum energy scale \cite{Copeland/2006,Nojiri/2007,Tsujikawa/2010}. Therefore, it is important to investigate alternative approaches or generalizations of fundamental theories of gravity.\

 %The modified gravity theory became a fundamental component of modern cosmology. The possibility is that a modification to the classical gravitational action may solve various cosmic issues, such as the inflationary paradigm, dark energy, and dark matter. It successfully solves the coincidence problem since dark matter and dark energy result from the universe's expansion,  which is guided by a specific theory. Various modified gravity models can effectively capture the gravitational influence of dark matter.
 
 The simplest generalization of GR is the so-called $f(R)$ gravity, which generalizes the Hilbert-Einstein action to a general function of the Ricci scalar $R$ \cite{Buchdahl/1970,Barrow/1983}. This modified theory of gravity is well-known for its successful representation of cosmic acceleration, and also it can reproduce the entire cosmological history, as well as the behavior of the cosmological constant ($\Lambda$) \cite{Nojiri/2006,Elizalde/2009,Dombriz/2006}. In order to recover the properties of $\Lambda$CDM through modified theories of gravity, there are several methodologies adopted in the literature. But, among them, the cosmological reconstruction schemes have a special place. For instance, an interesting scheme for the cosmological reconstruction of $f(R)$ gravity in terms of e-folding was developed by Nojiri et al. \cite{Nojiri/2009}. For an accurate reconstruction of $\Lambda$CDM evolution in $f (R)$ gravity, Dunsby et al. \cite{Dunsby/2010} determined that the matter components must have an extra degree of freedom.
 Carloni et al. \cite{Carloni/2012} reconstructed $f (R)$ gravity by utilizing cosmic parameters instead of any form of scale factors. Goheer et al. \cite{Goheer/2009a} demonstrated that exact power-law solutions in $f(G)$ gravity exist exclusively for a specific class of models because many popular $f(G)$ models ( all examples addressed in \cite{Felice/2009}) do not permit any exact power-law solutions. This result is also to that found for $f(R)$ and $f(R,T)$ gravity models \cite{Goheer/2009b,Sharif/2013}. In the context of $f(R, T)$ gravity, Jamil et al. \cite{Jamil/2012} have reconstructed cosmological models and demonstrated that the dust fluid reproduces the $\Lambda$CDM,  de Sitter Universe, Einstein static Universe, phantom and non-phantom era, and phantom cosmology. In \cite{Sharif/2014}, have represented the cosmological reconstruction of $f(R,T)$ models and their stability. An analysis of the stability of $f (R, G)$ models is presented in \cite{Dombriz/2012} for power law and $\Lambda$CDM cosmology. Other than these curvature-based modified theories of gravity, there are two types of modified theories of gravity by which one can explain the universe: teleparallel gravity and symmetric teleparallel gravity theories.
 
 In this work, we aim to study the more recent modifications based on a Lagrangian density that is a generic function of the non-metricity scalar $Q$ that has been investigated \cite{Jimenez/2018}. At the background level, the modified theory of $f(Q)$ gravity leads to fascinating cosmological phenomenology, one can see \cite{Jimenez/2020,Dialektopoulos/2019,Bajardi/2020,Mandal/2020a,Mandal/2020b,Arora/2022}. Several studies in the literature indicate that the $f(Q)$ theory is one of the favorable alternative gravity formulations for explaining cosmological observations \cite{Valentino/2021,Yang/2020,Yang/2021}. A class of $f(Q)$ theories with Q coupled non-minimally to the matter Lagrangian was developed by Harko et al. \cite{Harko/2018}. They demonstrate through a cosmic application that it can serve as an alternate approach for understanding dark energy (DE). Testing against a variety of recent background data, including Type Ia Supernovae, Pantheon data, Hubble data, etc., has been done to provide observational limits on the background behavior of numerous $f(Q)$ models \cite{Lazkoz/2019,Ayuso/2021}. Furthermore, $f(Q)$ gravity comfortably passes the constraints imposed by Big Bang Nucleosynthesis (BBN) \cite{Anagnostopoulos/2020}. Using reconstruction methods, F. Esposito et al. \cite{Esposito/2022} investigated exact isotropic and anisotropic cosmological solutions.
 
 Inspired by the fascinating characteristics of $f(Q)$ gravity, we aim to develop a class of $f(Q)$ functions, which will be able to mimic the properties of the $\Lambda$CDM model. For this purpose, we adopt a number of explicit reconstructions, which lead to a number of interesting results categorized in two schemes. In the first scheme, we obtain some real values $f(Q)$ functions that are able to retrieve the $\Lambda$CDM expansion history of the universe filled with various matter components, respectively. In the second scheme, in terms of e-folding, a cosmological reconstruction for modified $f(Q)$ gravity is constructed. It is shown how any FLRW cosmology can arise from a specific $f(Q)$ theory. We further show that a theory can be constructed that can mimic the $\Lambda$CDM expansion history, which can be impossible to distinguish from GR at the fundamental level of the FLRW universe.\\
 
 This manuscript is arranged as follows. A general action and motion equation for $f(Q)$ gravity is presented in section \ref{section 2}. In section \ref{section 3}, we reconstruct the cosmological models for the various kind of fluid. In section \ref{section 4}, in terms of e-folding, we reconstruct the cosmological models for modified $f(Q)$ gravity. Lastly, discussion and conclusions are presented in Section \ref{section 5}.

\section{Basic Formalism of $f(Q)$ gravity}
\label{section 2}
Consider the most general $f(Q)$ gravity action, which is given by \cite{Jimenez/2018}
\begin{equation}
\label{1}
S=\int \left[\frac{1}{2}f(Q)+\mathcal{L}_m\right]\sqrt{-g}d^4x,
\end{equation}
where $Q$ is the non-metricity scalar, $f$ is an arbitrary function of $Q$,
 $\mathcal{L}_m$ is the matter Lagrangian density, and $g$ is a determinant of the metric tensor $g_{\alpha\beta}$. Here, we assume $8\pi G =1$. $Q_{\sigma\alpha\beta}=\nabla_{\sigma}\,g_{\alpha\beta}$ is the definition of nonmetricity tensor in $f(Q)$ gravity and the corresponding  traces are 
\begin{equation}
\label{2}
Q_{\sigma}=Q_{\sigma\,\,\,\,\alpha}^{\,\,\,\,\alpha}\, ,\,\,\,\,\,\,\,\,\tilde{Q}_{\sigma}=Q^{\alpha}_{\,\,\,\,\sigma\alpha}\,.
\end{equation}
Moreover, the superpotential tensor $P_{\,\,\mu\nu}^{\lambda}$ is given by
\begin{equation}
\label{3}
4P_{\,\,\alpha\beta}^{\sigma}=-Q^{\sigma}_{\,\,\,\,\alpha\beta}+2Q^{\,\,\,\,\,\,\sigma}_{(\alpha\,\,\,\,\beta)}-Q^{\sigma}g_{\alpha\beta}-\tilde{Q}^{\sigma}g_{\alpha\beta}-\delta^{\sigma}_{(\alpha}\, Q\,_{\beta)},
\end{equation} 
acquiring the nonmetricity scalar as 
\begin{equation}
\label{4}
Q=-Q_{\sigma\alpha\beta}P^{\sigma\alpha\beta}.
\end{equation}
The energy-momentum tensor for matter read as
\begin{equation}
\label{5}
T_{\alpha\beta}\equiv-\frac{2}{\sqrt{-g}}\frac{\delta(\sqrt{-g})\mathcal{L}_m} {\delta g^{\alpha\beta}}.
\end{equation}
The gravitational field equation of $f(Q)$ gravity shown below is derived by varying the action \eqref{1} over $g_{\mu\nu}$:

\begin{multline}
\label{6}
\frac{2}{\sqrt{-g}}\nabla_{\sigma}\left(f_{Q}\sqrt{-g}\,P^{\sigma}_{\,\,\alpha\beta}\right)+\frac{1}{2}f\,g_{\alpha\beta}+\\
f_{Q}\left(P_{\alpha\sigma\lambda}Q_{\beta}^{\,\,\,\sigma\lambda}-2Q_{\sigma\lambda\alpha}P^{\sigma\lambda}_{\,\,\,\,\,\,\beta}\right)=- T_{\alpha\beta},
\end{multline}
where $f_Q=\frac{d f}{d Q}$. The equation \eqref{1} can also be varied with respect to the connection as follows:
\begin{equation}
\label{7}
\nabla_{\alpha}\nabla_{\beta} \left(f_{Q}\sqrt{-g}\,P_{\,\,\,\,\,\,\,\sigma}^{\alpha\beta}\right)=0.
\end{equation}

Throughout the study, we will assume a spatially flat FLRW Universe, whose metric is given by 
\begin{equation}
\label{8}
ds^2=-dt^2+a^2(t)(dx^2+dy^2+dz^2).
\end{equation}
Here, $a(t)$ is a scale factor and $t$ is the cosmic time. The non-metricity scalar in FLRW metric is obtained as $Q=6H^2$, where $H=\frac{\dot{a}}{a}$ is the Hubble function and the over-head dot indicates the derivative with respect to $t$. In this case, the energy-momentum tensor of a perfect fluid $T_{\alpha\beta}=(p+\rho)u_{\alpha}u_{\beta}+pg_{\alpha\beta}$, where $\rho$ and $p$ are energy density and pressure, respectively.\\ For the metric \eqref{8}, the corresponding Friedmann equations are \cite{Jimenez/2018}: 
\begin{equation}
\label{9}
3 H^2=\frac{1}{2f_Q}\left(-\rho+\frac{1}{2}f\right),
\end{equation} 
\begin{equation}
\label{10}
\dot{H}+3H^2+\frac{\dot{f}_Q}{f_Q}H=\frac{1}{2f_Q}\left(p+\frac{1}{2}f\right).
\end{equation}

Now, one can investigate various cosmological applications using above Friedmann equations in the background of $f(Q)$ gravity. Further, we would like to mention here that the $f(Q)$ gravity satisfies the conservation equation
\begin{equation*}
\dot{\rho}+3H(\rho+p)=0.
\end{equation*}

\section{Reconstruction of a $f(Q)$ theory that admits an exact $\Lambda$CDM model}
\label{section 3}
\label{3}
In this section, we reconstruct the $f(Q)$ gravity model for various epochs such that it can precisely mimic the $\Lambda$CDM model. We can develop a real-valued function for the non-metricity scalar that can provide the particular cosmological evolution of the $\Lambda$CDM model. The observational cosmology suggests that the Hubble rate in terms of redshift, which is described in the $\Lambda$CDM model, is given by
\begin{equation}
\label{11}
H(z)=\sqrt{\frac{\rho_0}{3}(1+z)^3+\frac{\Lambda}{3}},
\end{equation}
where $\Lambda$ is a cosmological constant and $\rho_0\geq 0$ is the matter density. In follows, we try to construct the $f(Q)$ gravity theories that exactly mimic the $\Lambda$CDM expansion.\\
Using the scale factor $a$ and redshift $z$ relationship as $\frac{1}{a}=1+z$, the above equation can be demonstrated as
 
\begin{equation}
\label{12}
\frac{\dot{a}}{a}=\sqrt{\frac{\rho_0}{3a^3}+\frac{\Lambda}{3}}.
\end{equation}
From the above equation, we may get the derivative of the scale factor $a(t)$ with respect to time ($t$) as
\begin{equation}
\label{13}
\dot{a}=\sqrt{\frac{\rho_0}{3a}+\frac{\Lambda}{3}a^2}.
\end{equation}
%We know that the non-metricity scalar for a flat FLRW universe is defined by
%\begin{equation}
%\label{14}
%Q=6H^2=6\left(\frac{\dot{a}}{a}\right)^2
%\end{equation}
Now, we can rewrite the non-metricity scalar $Q$ in terms of the scale factor by plugging Eq.\eqref{12},
\begin{equation}
\label{15}
Q(a)=\frac{2(\rho_0+\Lambda\, a^3)}{a^3}.
\end{equation}
With the help of above equation, we write the scale factor in terms of non-metricity scalar as 
\begin{equation}
\label{16}
a(Q)=\left(\frac{2\,\rho_0}{(Q-2\,\Lambda)}\right)^{\frac{1}{3}}.
\end{equation}
We would like to note here that so far, we have not presumed \textit{a priori} and a specific theory of gravity. Expression for Hubble parameter is obtained by observations, and the non-metricity scalar is completely geometrical for FLRW spacetime, independent of gravitational theory. Further, the scale factor in Eq. \eqref{16} has one real and two complex roots. The scale factor has to be real. Therefore, with the real value of the scale factor, the non-metricity scaler reaches the value $2\Lambda$ in an infinite time. From Eq.\eqref{16}, we can calculate the Hubble parameter in terms of non-metricity scalar as 
\begin{equation}
\label{17}
H(Q)=\frac{1}{a(Q)}\sqrt{\frac{\rho_0}{3a(Q)}+\frac{\Lambda}{3}\,a^2(Q)}.
\end{equation}
Now, in order to find a class of $f(Q)$ functions, which mimic the $\Lambda$CDM expansion, , we plug all of the above quantities represented as functions of the non-metricity scalar into the Friedmann equation Eq.\eqref{9}, yielding the first order inhomogeneous differential equation for the function $f(Q)$ in $Q$-space,
\begin{equation}
\label{18}
Q\,\frac{df(Q)}{dQ}+\rho(Q)-\frac{f(Q)}{2}=0.
\end{equation}
Utilizing the energy conservation equation, the inhomogeneous term $\rho(Q)$ may be derived in terms of nonmetricity $Q$. The inhomogeneous term disappears in the vacuum case, yielding the homogeneous first order differential equation, and its solution is $f(Q)=c\,\sqrt{Q}$, where $c$ is a integration constant.\\ 
In further study, we only focus on the the inhomogeneous term $\rho(Q)$ to construct more theories for the $\Lambda$CDM expansion. let us explicitly reconstruct the theories that would lead to a particular matter field obeying a $\Lambda$CDM expansion history.  

\subsection{Reconstruction for dust-like matter}
First, we reconstruct the $f(Q)$ model, which may represent the $\Lambda$CDM era while taking dust into account as a matter content. For the dust like case, the value equation of state $w=0$. Imposing the dust like case in the conservation equation, we obtained the energy density in terms of $a(t)$ as
\begin{equation}
\label{19}
\rho(a)=\frac{\rho_0}{a^3}.
\end{equation}
With the help of Eq.\eqref{16}, rewrite above equation as
\begin{equation}
\label{20}
\rho(Q)=\frac{Q-2\Lambda}{2}.
\end{equation}
Substitute the above value in differential Eq.\eqref{18}, we get the general solution as
\begin{equation}
\label{21}
f(Q)=-Q-2\,\Lambda+c_1 \sqrt{Q},
\end{equation}
where $c_1$ is a integration constant.\\
This finding is intriguing because it demonstrates that the only real-valued Lagrangian $f(Q)$ capable of accurately simulating the $\Lambda$CDM expansion history for a universe composed primarily of dust-like stuff. Also if we put $\Lambda=0$ in Eq. \eqref{21}, the function for $f(Q)$ recover the dust-dominated universe other than GR.

\subsection{Reconstruction for perfect fluid with EoS $p=-\frac{1}{3}\rho$}
In this scenario, we reconstruct the $f(Q)$ model in which the universe is accelerating, and the EoS value $w=-\frac{1}{3}$ is physically interesting because it sits near the limit of the set of matter fields that obey the strong energy condition.
Imposing the value of EoS $p=-\frac{1}{3}\rho$ in conservation equation, we obtained the energy density in terms of $a(t)$ as
\begin{equation}
\label{22}
\rho(a)=\frac{\rho_0}{a^2}.
\end{equation}
With the help of Eq.\eqref{16}, rewrite above equation as
\begin{equation}
\label{23}
\rho(Q)=\left(\frac{\rho_0^2(Q-2\Lambda)}{2}\right)^{\frac{2}{3}}.
\end{equation}

Substitute the above value in differential Eq.\eqref{18}, we get the general solution as
\begin{equation}
\label{24}
f(Q)=c_1\sqrt{Q}+\mu_1\left(\frac{Q-2\Lambda}{2\Lambda-Q}\right)^{\frac{2}{3}}\times _2 F_1\left(-\frac{2}{3},-\frac{1}{2},\frac{1}{2};\frac{Q}{2\Lambda}\right),
\end{equation}
where $_2 F_1$ is a Hypergeometric function and $\mu_1=2\left(\frac{\rho_0^2}{\Lambda}\right)^{\frac{2}{3}}$.

\subsection{Reconstruction for Multifluids}
Our present knowledge of the Universe is based on the interactions of multiple "matter" components, which primarily interact via gravity and electromagnetic radiation. The nature of the various components and possible interactions is often based on the concept of coupled ideal fluids. to shed some light in this type of scenario, we start from the same place as in the previous sections, but with two fluid species instead of one. Let us now assume that the universe contains both dust-like matter and a non-interacting stiff fluid, with current densities of $\rho_0$ and $\rho_s$, respectively. In this case, total matter density is calculated using the conservation equation,
\begin{equation}
\label{25}
\rho(a)=\frac{\rho_0}{a^3}+\frac{\rho_s}{a^6},
\end{equation} 
With the help of Eq.\eqref{16}, rewrite above equation as
\begin{equation}
\label{26}
\rho(Q)=\frac{Q-2\Lambda}{2}+\frac{\rho_s}{\rho_0^2}\frac{(Q-2\Lambda)^2}{4}.
\end{equation}
Substitute the above value in differential Eq.\eqref{18}, we get the general solution as
\begin{equation}
\label{27}
f(Q)=-Q-2\Lambda+\mu_1 Q-\mu_2 Q^2+\mu_3+c_1\sqrt{Q},
\end{equation}
where $\mu_1=\frac{2\rho_s\Lambda}{\rho_0^2}$, $\mu_2=\frac{\rho_s}{6\rho_0^2}$, $\mu_3=\frac{2\Lambda^2\rho_s}{\rho_0^2}$.\\
As a result, we conclude that the above-described theory of gravity would perfectly reproduce the history of the $\Lambda$CDM expansion if the universe is composed of minimally connected noninteracting massless scalar fields with dust like stuff.

\subsection{Reconstruction for nonisentropic perfect fluids}

The equation of state is used to characterise nonisentropic perfect fluids as
\begin{equation}
\label{28}
p=h(\rho,a).
\end{equation}
We may obtain the appropriate differential equation for $\rho$ by using the energy conservation relation,
\begin{equation}
\label{29}
\frac{d\rho}{da}=-\frac{3}{a}(\rho+h(\rho,a)).
\end{equation}
For much simplicity of above equation, we assume that the separable function of $h(\rho,a)$ as $h(\rho,a)=w(a)\rho$. In this scenario, we can simply integrate Eq.\eqref{29} to obtain
\begin{equation}
\label{30}
\rho(a)=Exp[-3\int\frac{1+w(a)}{a}\,da].
\end{equation}
Let us use an example where the time-dependent barotropic index is represented by
\begin{equation}
\label{31}
w(a)=\frac{2\alpha-\beta a^3}{\alpha+\beta a^3},
\end{equation}
where $\alpha$ and $\beta$ are constants.
Using Eq.\eqref{31} in Eq.\eqref{30}, we get
\begin{equation}
\label{32}
\rho(a)=\frac{(\alpha+\beta a^3)^3}{a^9}.
\end{equation}
With the help of Eq.\eqref{16}, rewrite above equation as
\begin{equation}
\label{33}
\rho(Q)=\frac{(2\rho_0\beta+\alpha Q-2\Lambda \alpha)^3}{8\rho_0^3}.
\end{equation}

Substitute the above value in differential Eq.\eqref{18}, we get the general solution as
\begin{equation}
\label{34}
f(Q)=-\mu_1Q+\mu_2Q^2-\mu_3Q^3-\mu_4+c_1\sqrt{Q},
\end{equation}
where, $\mu_1=\frac{3\alpha(\Lambda\alpha-\beta \rho_0)^2}{\rho_0^3}$, $\mu_2=\frac{\alpha^2(\Lambda\alpha-\beta \rho_0)}{2\rho_0^3}$, $\mu_3=\frac{\alpha^3}{20\rho_0^3}$ and $\mu_4=\frac{2(\Lambda\alpha-\beta\rho_0)^3}{\rho_0^3}$.\\
Let us take a look at another type of nonisentropic perfect fluid whose equation of state is
\begin{equation}
\label{35}
p=w\rho+h(a).
\end{equation}

Integrating energy conservation Eq.\eqref{29}, we obtained 
\begin{equation}
\label{36}
\rho(a)=\left(-\int 3a^{(2+3w)}h(a)\,da+c\right)a^{-3(1+w)}.
\end{equation}
Let us use the particular cases of $h(a)=a^{-12}$ and $w=0$. This suggests that the universe's matter field is made up of dust and a time-dependent cosmological component that diverges at the Big Bang singularity and quickly decays to zero at subsequent epochs.

After solving Eq.\eqref{36}  we obtained matter density as 
\begin{equation}
\label{37}
\rho(a)=\frac{\rho_0}{a^3}+\frac{\rho_1}{a^{12}}.
\end{equation}

With the help of Eq.\eqref{16}, rewrite above equation as
\begin{equation}
\label{38}
\rho(Q)=\frac{Q-2\Lambda}{2}+\frac{\rho_1}{\rho_0^4}\frac{(Q-2\Lambda)^4}{16}.
\end{equation}

Substitute the above value in differential Eq.\eqref{18}, we get the general solution as
\begin{equation}
\label{39}
f(Q)=-Q+\mu_1Q-\mu_2Q^2+\mu_3Q^3-\mu_4Q^4+\mu_5+c_1\sqrt{Q},
\end{equation}
where $\mu_1=\frac{4\Lambda^3\rho_1}{\rho_0^4}$, $\mu_2=\frac{\Lambda^2\rho_1}{\rho_0^4}$, $\mu_3=\frac{\Lambda\rho_1}{5\rho_0^4}$, $\mu_4=\frac{\rho_1}{56\rho_0^4}$ and $\mu_5=-2\Lambda+\frac{2\Lambda^4\rho_1}{\rho_0^4}$.

\section{Cosmological reconstruction of $f(Q)$ gravity}
\label{section 4}
The first Friedmann equation corresponding to the FLRW equation can be rewrite as,
\begin{equation}
\label{40}
0=\frac{f(Q)}{2}-6H^2f'(Q)-\rho,
\end{equation}
where $Q=6H^2$ and $f_Q=f'$.
The above Friedmann equations are written as functions of the number of e-foldings instead of the time $t$, $N=log\frac{a}{a_0}$. The variable $N$ is related with the redshift $z$ by $e^{-N}=\frac{a_0}{a}=(1+z)$.  By adding the fluid densities with a constant EoS parameter $w_i$, we may get the matter energy density $\rho$ as
\begin{equation}
\label{41}
\rho=\sum_i \rho_{i0}\, a^{-3(1+w_i)}=\sum_i \rho_{i0}\,a_0^{-3(1+w_i)} e^{-3(1+w_i)N}.
\end{equation}
Let us write the Hubble parameter in terms of $N$ via the function $g(N)$ as 
\begin{equation}
\label{42}
H=g(N)=g(-ln(1+z)).
\end{equation}
Then non-metricity scalar written as $Q=6H^2=6g(N)^2$, where $N=N(Q)$. 
By using Eq.\eqref{41} and Eq.\eqref{42}, Eq.\eqref{40} can be written as,
\begin{equation}
\label{43}
0=\frac{f(Q)}{2}-6G(N)\frac{df(Q)}{dQ}-\sum_i \rho_{i0}\,a_0^{-3(1+w_i)} e^{-3(1+w_i)N}.
\end{equation}
Here we denote $G(N)=g(N)^2=H^2$.
As an example, consider the CDM model. As will be demonstrated below, for the gravity theory represented by action \eqref{1}, such evolution may be reconstructed without the need of a cosmological constant factor, and the shift from a decelerated to an accelerated period is achieved. The FLRW equation for $\Lambda$CDM cosmology in Einstein gravity is given by
\begin{equation}
\label{44}
\frac{3}{\kappa^2}H^2=\frac{3}{\kappa^2}H_0^2+\frac{\rho_0}{a^3}=\frac{3}{\kappa^2}H_0^2+\rho_0\,a_0^{-3}e^{-3N},
\end{equation}
$H_0$ and $\rho_0$ are constants in this case. The cosmological constant is represented by the first component in the RHS, while cold dark matter is represented by the second term (CDM). In the present universe, the (effective) cosmological constant $\Lambda$ is given by $\Lambda=12H_0^2$. Then comes the
\begin{equation}
\label{45}
G(N)=H_0^2+\frac{\kappa^2}{3}\rho_0\,a_0^{-3}e^{-3N},
\end{equation}
and $Q=6G(N)=6H_0^2+2\kappa^2\rho_0\,a_0^{-3}e^{-3N}$, which can be solved for $N$ as follows:
\begin{equation}
\label{46}
N=-\frac{1}{3}\,ln\left(\frac{Q-6H_0^2}{2\kappa^2\rho_0\,a_0^{-3}}\right).
\end{equation}
With the help of above equation, Eq.\eqref{43} can be taken as
\begin{equation}
\label{47}
0=\frac{f(Q)}{2}-Q\frac{df(Q)}{dQ}-K (Q-6H_0^2)^{(1+w)}.
\end{equation}
where $K=\rho_0\left( \frac{1}{2\kappa^2\rho_0}\right)^{(1+w)}$.
Solution of above differential equation is 
\begin{multline}
\label{48}
f(Q)=c_1\sqrt{Q}-2K(6H_0^2)^{(1+w)}\left(\frac{Q-6H_0^2}{6H_0^2-Q}\right)^w \\
\times _2 F_1\left(\frac{-1}{2},-1-w,\frac{1}{2};\frac{Q}{6H_0^2}\right),
\end{multline}
where $_2 F_1$ is a Hypergeometric function.\\
As a result, we showed that modified $f(Q)$ gravity can characterise the $\Lambda$CDM era without introducing the effective cosmological constant.\\

As another example, in Einstein gravity, we reconstruct $f(Q)$ gravity by replicating the system with non-phantom matter and phantom matter, whose FLRW equation is provided by

\begin{equation}
\label{49}
\frac{3}{\kappa^2}H^2=\rho_p\,a^{-c}+\rho_q\,a^c,
\end{equation}
where $\rho_p$, $\rho_q$ and $c$ are positive constants. We can verify that the first component in the R.H.S. in this solution relates to a non-phantom fluid with an equation of state (EoS) $w=-1+\frac{c}{3}>-1$, whereas the second term has an equation of state (EoS) $w=-1-\frac{c}{3}<-1$, which relates to a phantom fluid.
Then since $G(N)=g(N)^2=H^2$, we find 
\begin{equation}
\label{50}
G(N)=G_p\,e^{-cN}+G_q\,e^{cN}.
\end{equation}
where $G_p$ and $G_q$ are constants. Then since $Q=6G(N)$,
\begin{equation}
\label{51}
e^{cN}=\frac{Q\pm \sqrt{Q^2-144G_pG_q}}{12G_q}.
\end{equation}
Then above equation can be solved for $N$ as follows:
\begin{equation}
\label{52}
N=\frac{1}{c}ln\left(\frac{Q\pm \sqrt{Q^2-144G_pG_q}}{12G_q}\right).
\end{equation}

We consider $c = 4$ case. In the case, the non-phantom matter corresponding to the first term in the RHS of Eq.\eqref{49} could be radiation with $w = \frac{1}{3}$. Then Eq.\eqref{43} in this case is given by
\begin{equation}
\label{53}
0=\frac{f(Q)}{2}-Q\frac{df(Q)}{dQ}-\frac{\rho_{r0}}{a_0^4}\left(\frac{Q\pm \sqrt{Q^2-144G_pG_q}}{12G_q}\right).
\end{equation}
Solution of above differential equation is 
\begin{multline}
\label{54}
f(Q)=-\mu_1Q+c_1\sqrt{Q}-\mu_2\sqrt{\frac{144G_pG_q-Q^2}{Q^2-144G_pG_q}}\\
\times _2 F_1\left(-\frac{1}{2},-\frac{1}{4},\frac{3}{4};\frac{Q^2}{144G_pG_q}\right).
\end{multline}
Where $_2 F_1$ is a hypergeometric function and $\mu_1=\frac{\rho_{r0}}{6a_0^4G_q}$, $\mu_2=\frac{2\rho_{r0}}{a_0^4}\sqrt{\frac{G_p}{G_q}}$ are constants.

\section{Discussion}
\label{section 5}
Several modified theories of gravity have been proposed to fill the hole in the GR. However, the $\Lambda$CDM model is very successful in describing the accelerated expansion; but it has a few shortcomings. Therefore, in this manuscript, we have looked at the modified theories of gravity that can mimic the exact $\Lambda$CDM expansion history of the universe. We have adopted several cosmological reconstruction techniques for $f(Q)$ gravity to carry out this task. In our first approach, we have presumed different fluid components and obtained a class of $f(Q)$ theories. We have also reconstructed the $\Lambda$CDM universe in terms of e-folding without using an auxiliary scalar in intermediate calculations. The great advantage of working on e-folding is that one can reconstruct any type of requested FLRW cosmology, such as the oscillating universe, transition from deceleration to phantom phase without future singularity, the subsequent transition from deceleration to phantom super-acceleration leading to Big Rip singularity, and $\Lambda$CDM epoch.

In our study, we have obtained a class of $f(Q)$ theories that exactly mimic $\Lambda$CDM expansion history, even if it is impossible to distinguish from the GR using measurements of the background cosmological parameters. Then, it is an exciting problem to see how the perturbations studies (such as growth factor, structure formations, or cosmological gravitational waves from GR) in these $f(Q)$ theories can be verified experimentally \cite{Jimenez/2020, pert,Anagnostopoulos/2021}. In fact, the $f(Q)$ theories presented in this manuscript can be tested through the solar system test, which rules out the Lagrangians to have a modified theory that works for both local and cosmological scales. Furthermore, we have restricted the reconstruction scheme to the flat FLRW cases, whereas in non-flat FLRW cases, one additional term for curvature will appear in the Hubble parameter, non-metricity scalar $Q$, and the Friedmann equations. As a result, a highly non-linear differential equation will arise for $f(Q)$, which is an open problem for readers. Hence, developing these types of modified theories adds a strong agreement in favor of inflation, dark matter, and dark energy in the context of a unified gravitational alternative theory.

\section*{Data Availability Statement}
There are no new data associated with this article.

\section*{Acknowledgments}
GNG acknowledges University Grants Commission (UGC), New Delhi, India for awarding Junior Research Fellowship (UGC-Ref. No.: 201610122060). S.M. acknowledges Department of Science \& Technology (DST), Govt. of India, New Delhi, for awarding INSPIRE Fellowship (File No. DST/INSPIRE Fellowship/2018/IF180676). We are very much grateful to the honorable referee and to the editor for the illuminating suggestions that have significantly improved our work in terms of research quality, and presentation.

\end{document}